\documentclass[fleqn,usenatbib]{mnras}

\usepackage{newtxtext,newtxmath}

\usepackage[T1]{fontenc}

\DeclareRobustCommand{\VAN}[3]{#2}
\let\VANthebibliography\thebibliography
\def\thebibliography{\DeclareRobustCommand{\VAN}[3]{##3}\VANthebibliography}

\usepackage{graphicx}	
\usepackage{amsmath}	
\usepackage{amssymb}	

\title[$V \sin{i}$ as SB2 indicator]{Detection of 2460 SB2 candidates in the LAMOST-MRS, using projected rotational velocities and a binary spectral model.}

\author[M. Kovalev et al.]{
Mikhail Kovalev$^{1,2,3,4}$\thanks{E-mail: mikhail.kovalev@ynao.ac.cn},
Xuefei Chen$^{1,2,5}$,
Zhanwen Han$^{1,2,5}$
\\
$^{1}$Yunnan Observatories, China Academy of Sciences, Kunming 650216, China\\
$^{2}$Key Laboratory for the Structure and Evolution of Celestial Objects, Chinese Academy of Sciences, Kunming 650011, China\\
$^{3}$Sternberg Astronomical Institute, Leninskie Gory, Moscow 119992, Russia\\
$^{4}$Max Planck Institute for Astronomy, D-69117 Heidelberg, Germany\\
$^{5}$Center for Astronomical Mega-Science, Chinese Academy of Sciences, 20A Datun Road, Chaoyang District, Beijing 100012, China\\
}

\date{Accepted 31-08-2022. Received 24-08-2022; in original form 15-07-2022}

\pubyear{2022}

\def\kms{\,{\rm km}\,{\rm s}^{-1}}

\def\feh{\hbox{[Fe/H]}}

\newcommand{\teff}{{T_{\rm eff}}}
\newcommand{\rv}{{\rm RV}}
\def\Vmic{V_{\rm mic}}

\def\vsini{V \sin{i}}
\def\logg{\log{\rm (g)}}
\def\snr{\hbox{S/N}}
\def\drv{\hbox{|$\Delta$ RV|}}
\def\imp{f_{\rm imp}}
\newcommand{\ha}{\hbox{H$\alpha$}}

\begin{document}
\label{firstpage}
\pagerange{\pageref{firstpage}--\pageref{lastpage}}
\maketitle

\begin{abstract}
We present a new method for the detection of double-lined spectroscopic binaries (SB2) using $\vsini$ values from spectral fits. The method is tested on synthetic and real spectra from LAMOST-MRS. It can reliably detect SB2 candidates for double-lined binaries with $\vsini_1+\vsini_2<300\kms$ if the radial velocity separation is large enough. Using this method, we detect 2460 SB2 candidates, 1410 of which are new discoveries. We confirm the correlation between the radial velocity separation estimated by the binary model and $\vsini_0$ estimated by the single star model using the selected sample. Additionally, our method finds one new SB2 candidate in open cluster M~11. 
\end{abstract}

\begin{keywords}
binaries : spectroscopic -- techniques : spectroscopic -- stars individual: J115307.93+353528.2, J005425.62+081141.3, J064726.39+223431.7, J065032.45+231030.2, J055923.95+303104.02, J065001.65+222127.7, J053818.60-091754.4. 
\end{keywords}



\section{Introduction}

Modern spectroscopic surveys provide a huge amount of stellar spectra. Traditionally, the main analysis of survey data is focused on single stars, while spectroscopic binaries (SBn, where n is a number of spectral components) should be removed and analysed separately. SB2s with non-zero radial velocity separation ($\drv=|\rv_1-\rv_2|$) of spectral components are usually detected by analysis of multiple peaks of the cross-correlation functions (CCF) \citep[][]{merle2017,li2021,kounkel}, although it is possible to detect an SB2 with $\drv=0$ using composite spectroscopic model \citep[][]{bardy2018,m11} if spectral components are significantly different. Machine learning techniques are also used to find SB2s \citep{traven20,zhangbo22}.
\par
\cite{m11} selected SB2 candidates in open cluster M~11 by visual inspection of the spectra, which showed a significant improvement of the fit, when using the composite spectral model. They found a clear correlation between the $\drv$ estimated by the binary model and the value of the projected rotational velocity $\vsini_0$, derived from the single-star model for SB2 systems, and proposed to use this correlation for SB2 identification. In this work we develop this idea further as a new method for SB2 detection, thereby finding one new SB2 candidate in the previous M~11 data set and successfully apply it to detect 2460 SB2 candidates in the LAMOST (Large Sky Area Multi-Object fiber Spectroscopic Telescope) Medium Resolution Survey (MRS) \citep{lamostmrs}.
\par
The paper is organised as follows: in Sections~\ref{sec:obs} and \ref{sec:methods}, we describe the observations and methods. Section~\ref{results} presents our results. In Section~\ref{discus} we discuss the results. In Section~\ref{concl} we summarise the paper and draw conclusions.

\section{Observations}
\label{sec:obs}

LAMOST is a 4-meter quasi-meridian reflective Schmidt telescope with 4000 fibers installed on its $5\degr$ FoV focal plane. These configurations allow it to observe spectra for at most 4000 celestial objects simultaneously (\cite{2012RAA....12.1197C, 2012RAA....12..723Z}).
 For the analysis in this paper, we downloaded all available time-domain spectra from \url{www.lamost.org}.	We use the spectra taken at a resolving power of $R=\lambda/ \Delta \lambda \sim 7\,500$. Each spectrum is divided on two arms: blue from 4950\,\AA~to 5350\,\AA~and red from 6300\,\AA~to 6800\,\AA.~We convert the heliocentric wavelength scale in the observed spectra from vacuum to air using \texttt{PyAstronomy} \citep{pya}. Observations are carried out in MJD=58407.6-59567.8 days, spanning an interval of 1160 days.
 We selected only spectra stacked for whole night\footnote{ Each exposure contains a sequence of the several short 20 min individual exposures, which were stacked to increase $\snr$, see Appendix~\ref{sec:subepochs} with example of non-stacked spectra.} and apply a cut on the signal-to-noise ($\snr$). In total we have $592\,060$ spectra from $138\,114$ targets, where the $\snr\geq25$ ${\rm pix}^{-1}$ in any of spectral arms. The majority of the spectra (one half) sample the $\snr$ in the range of 40-250 ${\rm pix}^{-1}$. The number of exposures varies from 1 to 27 per target, as noisy exposures were not selected for many targets.
\par
For M~11 stars we use 265 infrared spectra from Gaia-ESO survey \citep{Gilmore2012}. They were observed with the GIRAFFE \citep{Pasquini2002} instrument on the VLT (Very Large Telescope) of the European Southern Observatory (ESO) using the HR21 setup ($\lambda=8475-8980$~\AA,~$R=16\,200$) and have $\snr$ in the range from 30 to 120 ${\rm pix}^{-1}$. 
For a detailed description of the M~11 dataset we refer the reader to \citet{m11}. 

\section{Methods} 
\label{sec:methods} 
\subsection{Spectroscopic analysis}
\label{sec:specfit}
The results for M~11 stars were taken from the previous work \citet{m11}.
\par
We use the same spectroscopic models and method as \cite{tyc} to analyse individual LAMOST-MRS spectra, see brief description below.
The normalised binary model spectrum is generated as a sum of the two Doppler-shifted normalised single-star spectral models ${f}_{\lambda,i}$\footnote{they are designed as a good representation of the LAMOST-MRS spectra, see Appendix~\ref{sec:payne} for details}, scaled according to the difference in luminosity, which is a function of the $\teff$ and stellar size. We assume both components to be spherical and use the following equation:    

\begin{align}
    {f}_{\lambda,{\rm binary}}=\frac{{f}_{\lambda,2} + k_\lambda {f}_{\lambda,1}}{1+k_\lambda},~
    k_\lambda= \frac{B_\lambda(\teff{_{,1}})~M_1}{B_\lambda(\teff{_{,2}})~M_2} 10^{\logg_2-\logg_1}
	\label{eq:bolzmann}
\end{align}
 where $k_\lambda$ is the luminosity ratio per wavelength unit, $B_\lambda$ is the black-body radiation (Plank function), $\teff$ is the effective temperature, $\logg$ is the surface gravity and $M$ is the mass. Throughout the paper we always assume the primary star to be brighter and define mass ratio as $q=M_1/M_2$, which is inverted in comparison with traditional definition $Q=1/q$.
\par
The binary model spectrum is later multiplied by the normalisation function, which is a linear combination of the first four Chebyshev polynomials \citep[similar to][]{Kovalev19}, defined separately for the blue and red arms of the spectrum. The resulting spectrum is compared with the observed one using \texttt{scipy.optimise.curve\_fit} function, which provides the optimal spectral parameters and radial velocities (RV) for each component plus the mass ratio $q={M_1}/{M_2}$ and two sets of four coefficients of the Chebyshev polynomials. We keep the metallicity equal for both components. In total we have 18 free parameters for a binary fit. We estimate the goodness of the fit parameter by reduced $\chi^2$.

\par 
Additionally, every spectrum is analysed by a single star model, which is identical to a binary model when both components have all equal parameters, so we fit only for 13 free parameters. Using this single star solution\footnote{throughout paper it has subscript ``0" or ``single"} we compute the difference in reduced $\chi^2$ between two solutions and the improvement factor ($\imp$), computed using Equation~\ref{eqn:f_imp} similar to \cite{bardy2018}. This improvement factor estimates the absolute value difference between two fits and weights it by the difference between the two solutions.

\begin{align}
\label{eqn:f_imp}
f_{{\rm imp}}=\frac{\sum\left[ \left(\left|{f}_{\lambda,{\rm single}}-{f}_{\lambda}\right|-\left|{f}_{\lambda,{\rm binary}}-{f}_{\lambda}\right|\right)/{\sigma}_{\lambda}\right] }{\sum\left[ \left|{f}_{\lambda,{\rm single}}-{f}_{\lambda,{\rm binary}}\right|/{\sigma}_{\lambda}\right] },
\end{align}
where ${f}_{\lambda}$ and ${\sigma}_{\lambda}$ are the observed flux and corresponding uncertainty, ${f}_{\lambda,{\rm single}}$ and ${f}_{\lambda,{\rm binary}}$ are the best-fit single-star and binary model spectra, and the sum is over all wavelength pixels.

\par
 The current paper focuses on SB2 identification of a single epoch spectrum, and we will present the results of the simultaneous analysis of the multiple spectra in our future paper (Kovalev et al. in prep).

\subsection{Selection of SB2 candidates using rotational velocities}
\label{sec:selection}
The logic of the selection is quite simple: we need to separate single stars from all binary solutions. If we try to fit the double-lined spectrum with $\drv>0$ using a single star model there are three possible outcomes:
\begin{enumerate}
    \item none of the spectral components are well constrained, but $\vsini_0$ proportional to $\drv$ is large enough that broadened model covers both of them;
    \label{case1}
    \item only one spectral component (usually the primary) is well fitted, while the other is completely ignored by the spectral model;
    \label{case3}
    \item the spectrum is poorly fitted as the synthetic model is unable to reproduce a real spectrum due to missing physics (usually emission lines, molecular bands etc.) or data processing artifacts.
    \label{case2}
\end{enumerate}
For all these cases binary model fits the spectrum much better and $\imp$ is big. However only case~\ref{case1} is useful to select SB2s. Even small $\drv$, which is insufficient to split spectral lines due to finite spectrograph's resolution,  will cause change to the spectral lines: they will become broader and the fitted value of $\vsini_0$ will increase in comparison with the value from the spectrum of that binary taken when $\drv=0$. Case~\ref{case3} can be excluded by requiring $\rv_{1}$ to differ from $\rv_0$, as such a case is useless for SB2 identification. However, it can be used, when several observations are available for a given target. Multiple epochs allow us to find an SB2 candidate by $\vsini_0$ variation, which it is proportional to $\drv$, although such a variation can be caused by a real change of $\vsini$. 

\par
Now we need to separate out the single stars. When a single star spectrum with $\vsini_0$ is fitted by a binary model there are three possible outcomes:
\begin{enumerate}
    \item the spectrum is well fitted by twin binary model, consisting of two components similar to the real star with small $\drv\sim0$,  therefore $\vsini_{1,2}\sim \vsini_0$ and $\vsini_1+\vsini_2\sim 2\vsini_0$;
    \label{case01}
    \item the spectrum is well fitted by the primary component, which is almost identical to the real star ($\vsini_1\sim\vsini_0$), and a small or negligible contribution from the secondary component, which can have any value $\vsini_2$, although usually it is quite large, so the secondary spectrum looks completely ``flat"; 
    \label{case02}
    \item the spectrum is poorly fitted as the synthetic model is unable to reproduce the real spectrum. In this case, the binary model is desperately trying to compensate for the missing spectral information by combining two spectral components, usually with significant improvement relative to best single star model.
    \label{case03}
\end{enumerate}
For cases~\ref{case01} and \ref{case02} $\imp$ is usually small, but for case~\ref{case03} it can be quite large.
Thus we can select SB2 candidates with $\drv>0$ by selecting the spectra with $\vsini_1+\vsini_2 +\vsini_{\rm min}< \vsini_0$ and $\rv_1\neq\rv_0$, while ``bad fits" from case~\ref{case03} can be excluded using a cut on $\imp$. $\vsini_{\rm min}$ takes into account the possible uncertainties in the $\vsini$ measurements. However we should note that these criteria will not select spectra of fast rotators, as in this case $\vsini_1+\vsini_2>\vsini_0$ for small $\drv>0$. They can be selected only for a significant $\drv$. One can find various fit examples in Appendix~\ref{fit_examples}.

\subsubsection{Test on synthetic SB2s}
\label{sec:err}
\begin{figure}
	\includegraphics[width=\columnwidth]{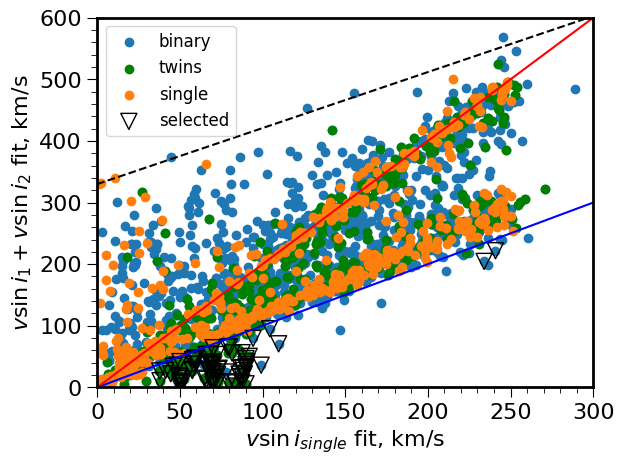}
	\includegraphics[width=\columnwidth]{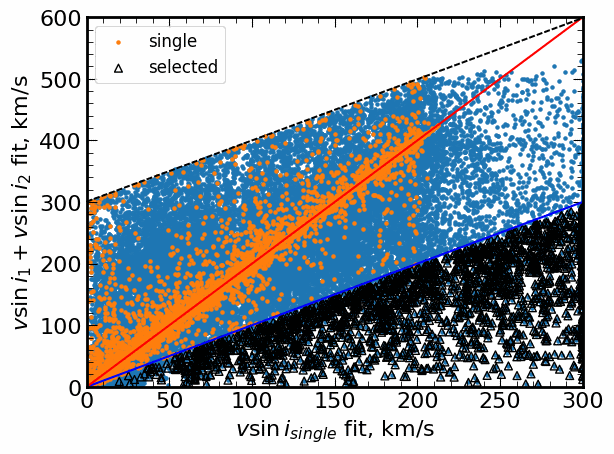}
    \caption{Example of the selection using $\vsini$ for simulated datasets of M~11 (top) and LAMOST-MRS (bottom). Solid lines show the functions $\vsini_1+\vsini_2=\vsini_0$ (blue) and $\vsini_1+\vsini_2= 2\vsini_0$ (red).  }
    \label{fig:m11sim}
\end{figure}

\citet{m11} presented spectral simulations for open cluster M~11 with 1480 SB2s and 480 single stars, where 480 SB2s are twin binaries, with all identical parameters for components except for $\rv$. Each star had only one spectrum. 
We take these results and plot $\vsini_1+\vsini_2$ versus $\vsini_0$ in the top panel of Fig.~\ref{fig:m11sim}. Solid lines show the functions $\vsini_1+\vsini_2=\vsini_0$ (blue) and $\vsini_1+\vsini_2= 2\vsini_0$ (red). It is clearly seen that the single star subset mostly follows the red line and there are no single stars below the blue line. Mock binary stars populate the entire parameter space between zero and the dashed line $\vsini_1+\vsini_2=330 + \vsini_0$. We select only datapoints $\vsini_{\rm min}=5~\kms$ below the blue line with $\imp>0.1$ and $|\rv_1-\rv_0|>5~\kms$ as SB2 candidates, to exclude all spectra from the single-star subset. In total we have 76 spectra, 59 of them are twin binaries, therefore success rate ($SR$) of our selection method is $SR=76/1480=0.05$ for all SB2s and $SR_{\rm twins}=59/480=0.125$ for twins. The success rate is small, because this simulated dataset mostly contains fast rotators with small $\drv$. 

Therefore for the time-domain LAMOST-MRS we create a new simulated set with larger $\drv$ and several spectra per star, using binary spectral model from Section~\ref{sec:specfit}.
We generate 10000 mock binaries using uniformly distributed mass-ratios $Q=M_2/M_1=U(0.01, 1.0)$, $\teff$ from $4600$ to $8800$ K, $\logg$ from 2.6 to 4.8 (cgs), $\vsini=U(1,200)\, \kms$ for both components and $\feh_{1}=\feh_{2}=U(-0.36,0.36)$ dex. For each star, we create five mock binary spectra using radial velocities computed for circular orbits with the random semiamplitudes $K_2=U(0,300), K_1=Q K_2\,\kms$ at randomly chosen phases. These models are degraded by Gaussian noise according to $\snr=50,\,100$ pix$^{-1}$ for the blue and red spectral arms respectively. Such models serve as a good representation of the real LAMOST-MRS observations.
We compute the secondary contribution to the total light at $\lambda=5000$~\AA~as $\mathrm{frac}=1/(1+k_{5000})$ and selected 1411 stars with $\mathrm{ frac}<0.01$ (below noise level) as a subset of the single stars with 7044 spectra.
\par
We perform exactly the same analysis as for the observations on this simulated dataset (see Section~\ref{sec:specfit}). We checked how well the parameters can be recovered by calculating the average and standard deviation of the residuals. For the single stars we have $\Delta\rv=0.03\pm0.32\kms$, $\Delta\teff_0=34\pm64$~K, $\Delta\logg_0=0.05\pm0.08$ cgs units, $\Delta\vsini_0=-1\pm2\kms$  and $\Delta\feh_0=0.02\pm0.04$ dex. For the primary components we have $\Delta\rv_1=-0.02\pm12.63\kms$, $\Delta\teff_1=11\pm192$~K, $\Delta\logg_1=0.05\pm0.14$ cgs units, $\Delta\vsini_1=0\pm13\kms$. For the secondary components we have $\Delta\rv_2=1.35\pm72.58\kms$, $\Delta\teff_2=157\pm891$~K, $\Delta\logg_2=0.29\pm0.55$ cgs units, $\Delta\vsini_2=-23\pm60\kms$. Metallicity has $\Delta\feh_{1,2}=0.01\pm0.04$ dex. The mass ratio is recovered poorly $\Delta Q=-0.05\pm0.33$, and thus is mostly unreliable. It is clear that the parameters of the secondary components are poorly recovered compared to the primary components, especially $\logg_{2}$. However accurate spectral parameters are not critical for the purpose of SB2 identification, as binary model serves as a ``flexible template". The simultaneous analysis of multiple epochs have a much better impact on a parameter's recovery\citep[][]{tyc}.
\par
We show the selection of SB2 candidates using $\vsini$ on bottom panel of Fig.~\ref{fig:m11sim}. Solid lines show the functions $\vsini_1+\vsini_2=\vsini_0$ (blue) and $\vsini_1+\vsini_2= 2\vsini_0$ (red). It is clearly seen that the single star subset mostly follows the red line and no single stars are seen below the blue line. Mock binary stars populate all parameter space between zero and dashed line $\vsini_1+\vsini_2=300 + \vsini_0$. We select all datapoints $\vsini_{\rm min}=5~\kms$ below the blue line with $\imp>0.1$ and $|\rv_1-\rv_0|>10~\kms$ as SB2 candidates. In total we have 2127 spectra of 1112 stars (zero from the single star subset), where all five spectra are selected for 22, four for 74, three for 189, two for 327 and only one for 500 mock binaries respectively. Therefore the success rate of our selection method is around $SR=1112/(10000-1411)=0.129$.   
\par
In our selection we prioritize purity over completeness, thus $SR$ is relatively small. One can use less strict quality cuts on $\vsini_{\rm min}$,\,$\imp$ and $|\rv_1-\rv_0|_{\rm min}$, however it can produce many false-positive detections. If multiple epochs are available one can find more SB2 candidates using $\vsini_0$ variation with time. We plan to implement this in improved selection method in our future paper (Kovalev et al. in prep.).

\section{Results}
\label{results}

\subsection{Application to binaries from M~11}
\begin{figure}
	\includegraphics[width=\columnwidth]{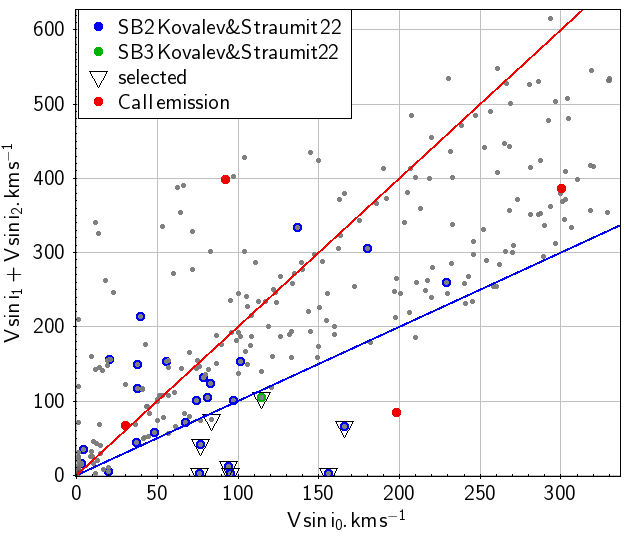}
	\includegraphics[width=\columnwidth]{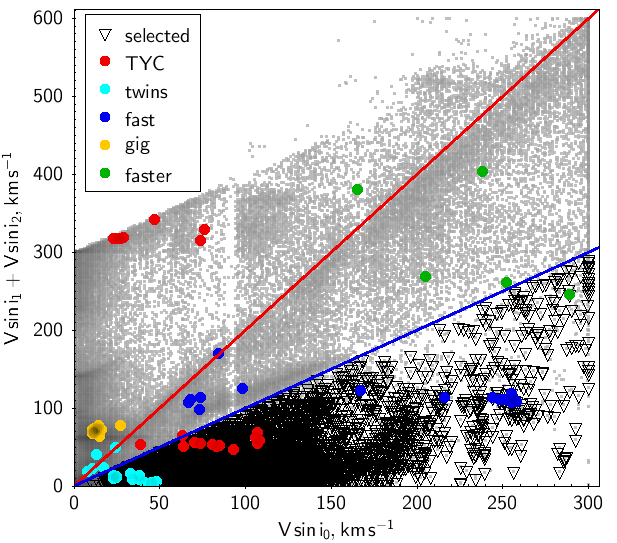}
    \caption{Example of the SB2 selection using $\vsini$ for observed datasets of M~11 (top panel) and LAMOST-MRS (bottom panel). Solid lines show the functions $\vsini_1+\vsini_2=\vsini_0$ (blue) and $\vsini_1+\vsini_2= 2\vsini_0$ (red). Selected stars are shown with black open triangles. Several confirmed SB2s are highlighted. }
    \label{fig:select}
\end{figure}

We use our criteria to explore the data from \citet{m11}. We show $\vsini$ plot in the top panel of Fig.~\ref{fig:select}. Four spectra with strong emission at Ca II lines are shown as red circles, they are poorly fitted based on visual inspection. We find that the new criteria select eight SB2 candidates out of 265 spectra, shown as black open triangles. All SB2s and SB3 candidates from \citet{m11} are shown as blue and green circles, where seven are selected, thus we find one new SB2 candidate. We discuss them in Appendix~\ref{sec:list_m11}.

\subsection{LAMOST-MRS}

\subsubsection{Quality cuts}
We carefully check the quality of the spectral fits through visual inspection of the plots.
Our LAMOST-MRS dataset contains spectra from various targets and some of them can be poorly fitted by our spectral model (red super giants, very hot stars, see Fig.~\ref{fig:bad_fits}). Therefore we introduce several quality cuts on the fitted parameters, see Table~\ref{tab:cuts}. They exclude $192\,533$ spectra (33 per cent) from the data set. Additionally we find 213 spectra, where wavelength scale is clearly shifted between the blue and left arms, see Appendix~\ref{sec:bads}.    

\begin{table}
    \centering
    \caption{Quality cuts}
    \begin{tabular}{l}
\hline
\hline
Filtering bad fits - any of the following cuts\\
\hline
$\chi^2_{\rm single}$  >120\\
$\drv$  >320 $\kms$\\
|$\rv_0$|  >400 $\kms$\\
|$\log{q}$|  >0.95\\
$k_{5000}/(1+k_{5000})$  >0.95\\
|$\feh_0$|>0.85 and $\vsini_0$>299 $\kms$\\
|$\logg_0-3$|>1.98 and $\vsini_0$>299 $\kms$\\
|$\logg_0-3$|>1.98 and ${\teff}_0$>8780 K\\
|$\logg_1-3$|>1.98 and ${\teff}_1$>8780 K\\
Bad $\lambda$ calibration in 213 spectra\\
\hline

    \end{tabular}
    \label{tab:cuts}
\end{table}

\subsubsection{Selected SB2 candidates}
We show the selection of SB2 candidates using $\vsini$ on the bottom panel of Fig.~\ref{fig:select}. It is clearly seen that not selected stars slightly follow red and blue lines. We have a large overdensity at $\vsini_0=1\,\kms$ and there is a clear gap at $\vsini_0\sim90\,\kms$.  We currently do not have a clear explanation for this gap, perhaps caused by a systematic problem in the fitting mechanism, since there is weak evidence of a similar gap for the simulated LAMOST-MRS dataset shown in Fig.~\ref{fig:m11sim}. Selected SB2s candidates are shown by open black triangles, with the majority having $\vsini_0$ in the interval from 20 to 150 $\kms$. In total we have 6160 spectra from 2460 stars, listed in Table~\ref{tab:final}. For several confirmed SB2s we highlight all available spectra as filled circles. It is clearly seen that many such spectra follow horizontal lines of nearly constant $\vsini_1+\vsini_2$. For example the Algol-like system TYC 2990-127-1 ($q=4.75$) from \citet{tyc} is labeled as ``TYC" (red circles). Many spectra are selected, although several spectra taken near conjunction phase are fitted with very big $\vsini_2$. Twin binary (cyan circles) with unresolved double-lined spectrum is selected many times, but some spectra with $\drv\sim0$ follow the red line. Two systems of fast rotators, labeled as ``fast" and ``faster" are shown as blue and green circles. For ``fast" fits of several spectra, where single star model was ignoring secondary component (case \ref{case3}), are not selected. In the ``faster" only one spectrum is selected. Fits for these three SB2s are shown in Fig.~\ref{fig:faster}. We also show as gold circles an interesting SB2 system with very different spectral components ($\Delta \teff>3000$ K) labeled as ``gig". None of its spectra are selected due to small $\drv$ and fast rotation of the hotter component, see Fig.~\ref{fig:gig} . This star can be selected based on big $\imp>0.3$. We will present a detailed study of it in our future paper.

\section{Discussion}
\label{discus}
We plot $\drv$ and $\vsini_0$ for all selected spectra and confirm the correlation between these parameters. 
We fit a straight line through the data points using $\imp$ as a weight and find: 
\begin{align}
\drv=0.86\, \vsini_0 + 3.90, R=0.98,  
\end{align}
where $R$ is a Pearson's product moment correlation coefficient. Since we have $\vsini_{\rm min}=5~\kms$ as our selection criteria, in theory we can select SB2s with $\drv=8.2~\kms$ based on this fit. 
The minimal $\drv\sim20\,\kms$, among selected SB2s, which is much smaller than $\drv=50~\kms$, required by CCF based methods developed for LAMOST-MRS spectra \citep[][]{li2021}, therefore our method should be able to find more SB2 candidates. 
However, many SB2s are still not detected, especially when the single star model fits only one component (case \ref{case3}) or the spectrum was observed at moment of small $\drv$, see Appendix~\ref{sec:miss}.  
\begin{figure}
	\includegraphics[width=\columnwidth]{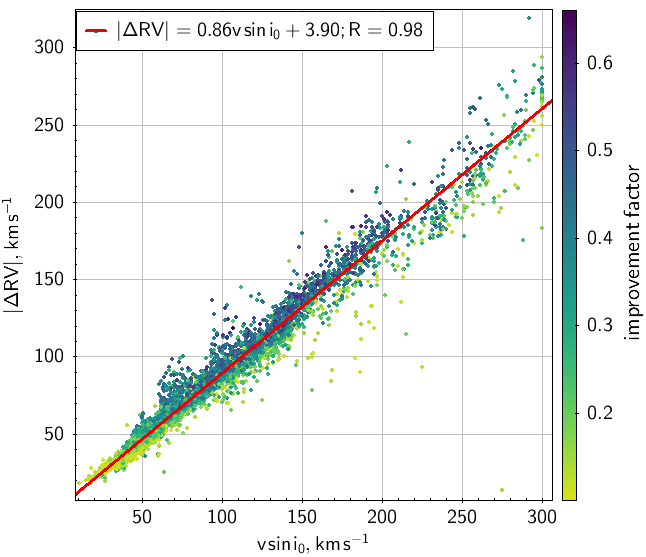}
    \caption{Correlation between $\drv$ and $\vsini_0$ for selected SB2s. }
    \label{fig:correlation}
\end{figure}

\par
We check how usage of stacked spectra can alter the fitting results in Appendix~\ref{sec:subepochs}. We find that for close SB2 with period $P<2$ days, stacking of 5 consecutive 20~min exposures leads to slight increase of $\vsini_{1,2}$ by $3-5~\kms$ in stacked spectrum, due to blurring. Therefore short epochs are more useful for close SB2s.    

\subsection{Comparison with other SB2 catalogues}
We make cross matches with several available SB2 catalogues. We find 17 matches with SIMBAD database \citep[][]{simbad}, labeled as SB*, two matches with SB9 \citep[][]{sb9},  56 matches with \cite{traven20}, and 108 matches with \cite{kounkel} catalogues. Cross matching with \cite{bardy2018} we find 10 stars listed in their SB2 table. Surprisingly, we find one match with their single star table and one match with the SB1 table. We check these matches and confirm that they are SB2 candidates, see Fig.\ref{fig:elbadry}. We find no matches with either the SB3 table or the SB2 table with unseen third component. Among LAMOST-MRS based papers we find 535 SB2 matches and 24 SB3 matches with \citet{li2021}, 173 matches with \citet{songK2} and 440 matches with \citet{zhangbo22}\footnote{among their final sample of 2198 SB2 candidates}. We find 122 matches with recent Gaia DR3 non-single stars orbits catalogue\citep[][]{gaia_dr3multiple}. 
\par
In total our catalogue includes 1050 known and 1410 new SB2 candidates. Also some of them are possibly SB3s, see Fig.\ref{fig:sb3}. However we should note that some SB2 candidates can be chance alignments due to relatively large fiber diameter of the spectrograph (3\arcsec). Our sample is limited to time-domain spectra with $\snr>25$, therefore the number of matches with other LAMOST-MRS based studies is not very high. We plan to extend it to non time-domain spectra in our future paper.

\subsection{Gaia DR3 data for LAMOST-MRS dataset}
\begin{figure}
	\includegraphics[width=\columnwidth]{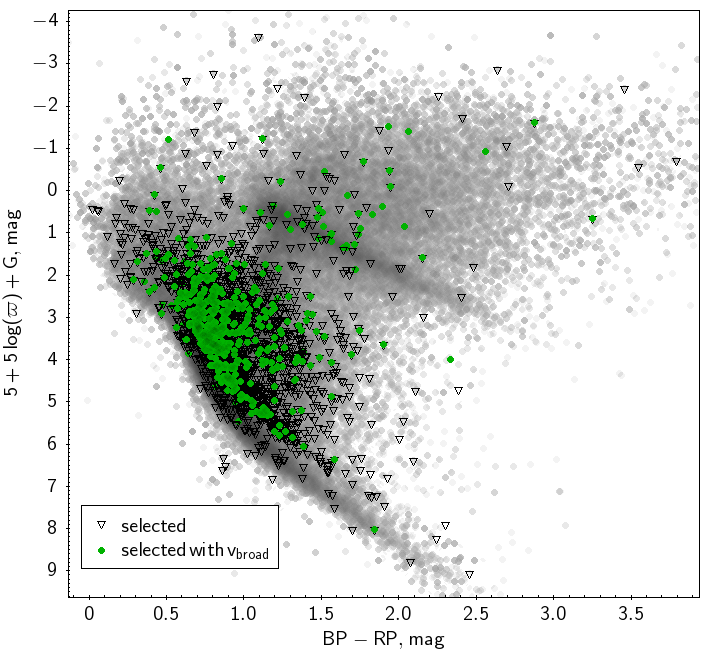}
    \includegraphics[width=\columnwidth]{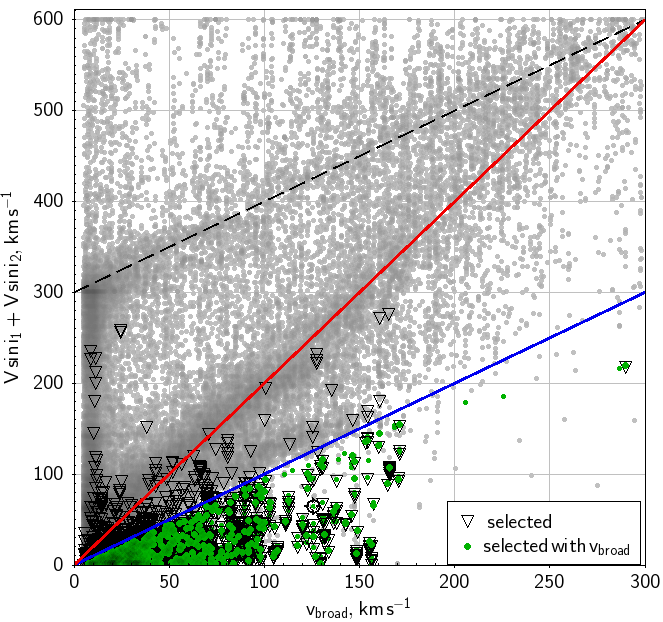}
    \caption{Gaia DR3 Hertzsprung-Russell diagram for observed LAMOST-MRS dataset (top panel) and $\vsini_1+\vsini_2$ versus $v_{\rm broad}$ (bottom panel). SB2 candidates selected in Fig.\protect\ref{fig:select} are shown with open black triangles, SB2 candidates selected using $v_{\rm broad}$ are shown with green circles.}
    \label{fig:hrd}
\end{figure}

We check the resent Gaia DR3 \citep[][]{gaia_all,gaia3} and plot the Hertzsprung-Russell diagram for all matches with positive parallax in top panel of Fig~\ref{fig:hrd}. 
Selected SB2 candidates are shown with black open triangles. Some of them are located at the main sequence of binaries with similar luminosity which is $\sim0.75$ mag higher than main sequence. Gaia DR3 provides parameter $v_{\rm broad}$ as a measure of rotational broadening for a single stars, based on Gaia RVS spectra \citep[][]{gaia_vbroad}. We use it instead of $\vsini_0$ and reproduce bottom panel of Fig.~\ref{fig:select} in the bottom panel of Fig.~\ref{fig:hrd}. Again is clearly seen that not selected stars slightly follow the red and blue lines. We have an overdensity at $v_{\rm broad}<10\,\kms$, and we have many hot stars in the region higher than the black dashed line $\vsini_1+\vsini_2=v_{\rm broad}+300$. In Fig.~\ref{fig:select} it was empty, because our fitting algorithm is unable fit $\teff>8800$~K, so it tries to increase $\vsini_0$ in order to compensate for changes in the spectra. We apply the same selection as before using $v_{\rm broad}$ instead of $\vsini_0$ and find 1380 spectra of 491 stars, where 436 of them were previously selected using $\vsini_0$. We show them as green circles. The remaining 55 stars were probably observed by LAMOST-MRS at moments with $\drv\sim0~\kms$. Unfortunately we can't check this hypothesis, because epoch's RVS spectra and RV measurements aren't included in Gaia DR3. However this confirms that \cite{gaia_vbroad} were unable to completely filter out SB2s from their catalogue. 

\section{Conclusions}
\label{concl}
We developed a new method for SB2 detection in spectral surveys. It is based on the simple fact that single star model will fit the SB2 spectrum with large rotational broadening $\vsini_0$, proportional to the $\drv$. This method found eight (one new) SB2 candidates in the M~11 cluster and 2460 (1410 new) SB2 candidates in the LAMOST-MRS. We will present a detailed study of the spectral parameters and orbits (similar to one in \cite{tyc}) of LAMOST-MRS SB2 candidates in our future paper. We hope that our method will be useful in SB2 detection in large-scale spectroscopic surveys, e.g. Gaia RVS \citep{grvs}.  
 

\section*{Acknowledgements}
 We are grateful to the anonymous referee for a constructive report. We thank Hans B{\"a}hr for his careful proof-reading of the manuscript.
MK is grateful to his parents, Yuri Kovalev and Yulia Kovaleva, for their full support in making this research possible. MK thanks Maria Kovaleva for valuable discussions.  This work is supported by National Key R\&D Program of China (Grant No. 2021YFA1600401/3), and by the Natural Science Foundation of China (Nos. 12090040/3, 12125303, 11733008).
Guoshoujing Telescope (the Large Sky Area Multi-Object Fiber Spectroscopic Telescope LAMOST) is a National Major Scientific Project built by the Chinese Academy of Sciences. Funding for the project has been provided by the National Development and Reform Commission. LAMOST is operated and managed by the National Astronomical Observatories, Chinese Academy of Sciences. The authors gratefully acknowledge the “PHOENIX Supercomputing Platform” jointly operated by the Binary Population Synthesis Group and the Stellar Astrophysics Group at Yunnan Observatories, Chinese Academy of Sciences. 
This work has made use of data from the European Space Agency (ESA) mission
{\it Gaia} (\url{https://www.cosmos.esa.int/gaia}), processed by the {\it Gaia}
Data Processing and Analysis Consortium (DPAC,
\url{https://www.cosmos.esa.int/web/gaia/dpac/consortium}). Funding for the DPAC
has been provided by national institutions, in particular the institutions
participating in the {\it Gaia} Multilateral Agreement.
Based on data products from observations made with ESO Telescopes at the La Silla Paranal Observatory under run IDs 188.B-3002 and 193.B-0936.
This research has made use of NASA’s Astrophysics Data System, the SIMBAD data base, and the VizieR catalogue access tool, operated at CDS, Strasbourg, France. It also made use of TOPCAT, an interactive graphical viewer and editor for tabular data \citep[][]{topcat}.

\section*{Data Availability}
The data underlying this article will be shared on reasonable request to the corresponding author.
LAMOST-MRS spectra are downloaded from \url{www.lamost.org}.
The M~11 data are based on public spectra downloaded from \url{http://archive.eso.org/wdb/wdb/adp/phase3_spectral/form?collection_name=GAIAESO}




\bibliographystyle{mnras}



\appendix

\section{Spectral model for LAMOST-MRS}
\label{sec:payne}
The grid of synthetic spectra (6200 in total) is generated using the NLTE~MPIA online-interface \url{https://nlte.mpia.de} \citep[see Chapter~4 in][]{disser} on wavelength intervals 4870:5430 \AA~for the blue arm and 6200:6900 \AA ~for the red arm with spectral resolution $R=7500$. We use a NLTE (non-local thermodynamic equilibrium) spectral synthesis for H, Mg~I, Si~I, Ca~I, Ti~I, Fe~I and Fe~II lines \citep[see Chapter~4 in][ for references]{disser}.  
The spectral parameters are randomly selected in a range of $\teff$=4600, 8800 K, $\logg$=1.0, 4.99 (cgs units), $\vsini$= 1, 300 $\kms$ and [Fe/H]\footnote{We used $\feh$ as a proxy of overall metallicity, abundances for all elements are scaled with Fe.}=$-$0.9,$+$0.9 dex,
microturbulence is fixed to $\Vmic=2~\kms$. The grid is randomly split on training (70\%) and cross-validation (30\%) sets of spectra, which are used to train \textit{The~Payne} spectral model \citep{ting2019}. 
We use output of \textit{The Payne} as a single-star spectral model ${f}_{\lambda,{\rm single}}$.

\section{Spectral fitting examples}
\label{fit_examples}
\subsection{Good fits}
\begin{figure*}
	\includegraphics[width=0.83\textwidth]{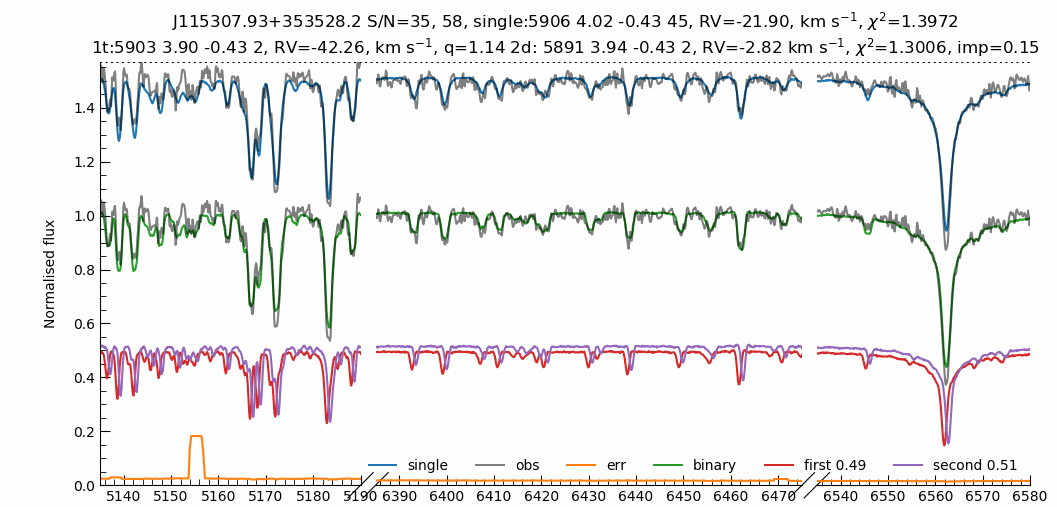}
	\includegraphics[width=0.83\textwidth]{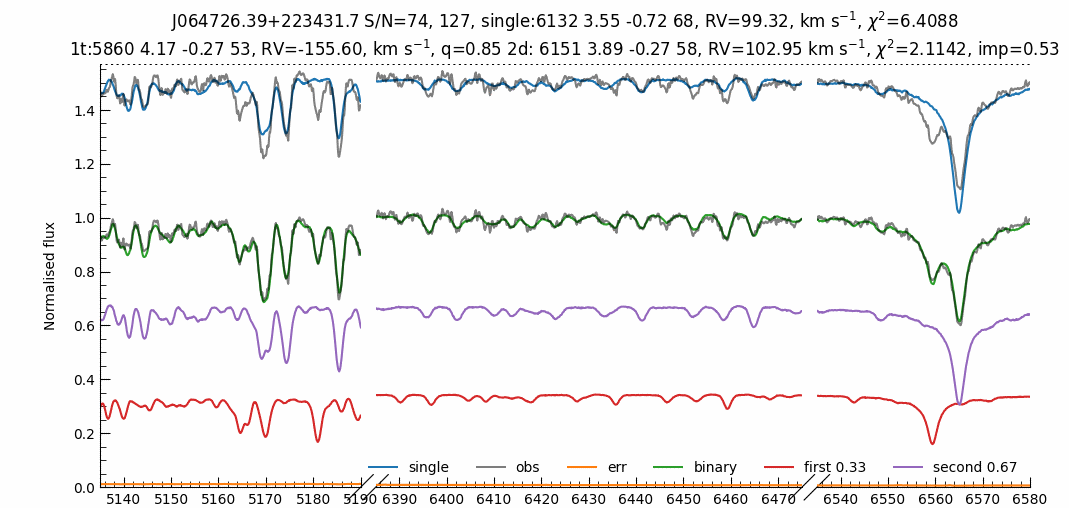}
	\includegraphics[width=0.83\textwidth]{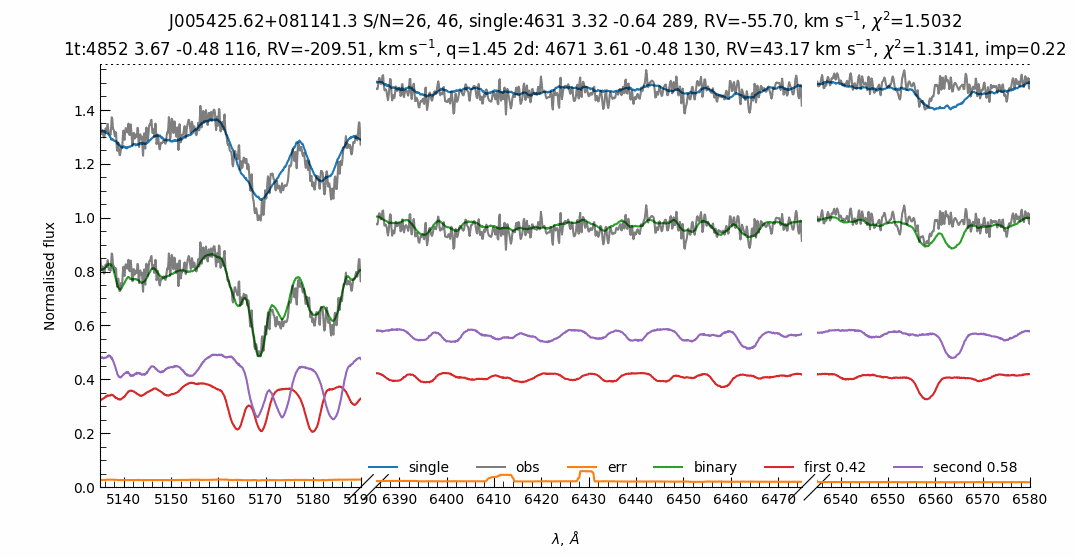}
    \caption{Examples of the spectrum fitting by binary spectral model (green line) and by single star model (blue line) (with offset 0.5). $\vsini_{1,2}$ are increasing from the top to bottom panel. Observed spectrum and its error are shown as a gray and orange lines respectively. Primary (magenta line) and secondary (red line) components are labeled as "second" and "first" with contribution to total light at $\lambda=5000$~\AA.~Spectral parameters ($\teff,~\logg,~\feh,~\vsini$) from single star model fit and binary model fit are shown in the titles.}
    \label{fig:faster}
\end{figure*}

In Figure~\ref{fig:faster} we show the best fit single star and binary models for several SB2s, confirmed by changes in spectral lines over several epochs. We zoom into the wavelength range around the magnesium triplet and $\ha$ and in a 70~\AA~interval in the red arm, where many double lines are clearly visible. In the top panel we show best fit of the twin binary spectrum of J115307.93+353528.2 (denoted as ``twins" in Fig.~\ref{fig:select}) taken at maximal $\drv=40~\kms$, however despite double-lines are not visible, this star is selected as binary by our method. Middle panel show an example of case \ref{case3} of SB2 fitting by the single star model, not useful for SB2 identification, however the binary model fits the spectrum well. This binary system J064726.39+223431.7 denoted as ``fast" in Fig.~\ref{fig:select}. In the bottom panel we show the best fit of the binary spectrum of system J005425.62+081141.3 (denoted as ``faster" in Fig.~\ref{fig:select}) of two fast rotators taken at significant $\drv$. We will study these SB2s in Kovalev et al. (in prep.).  

\begin{figure*}
	\includegraphics[width=0.83\textwidth]{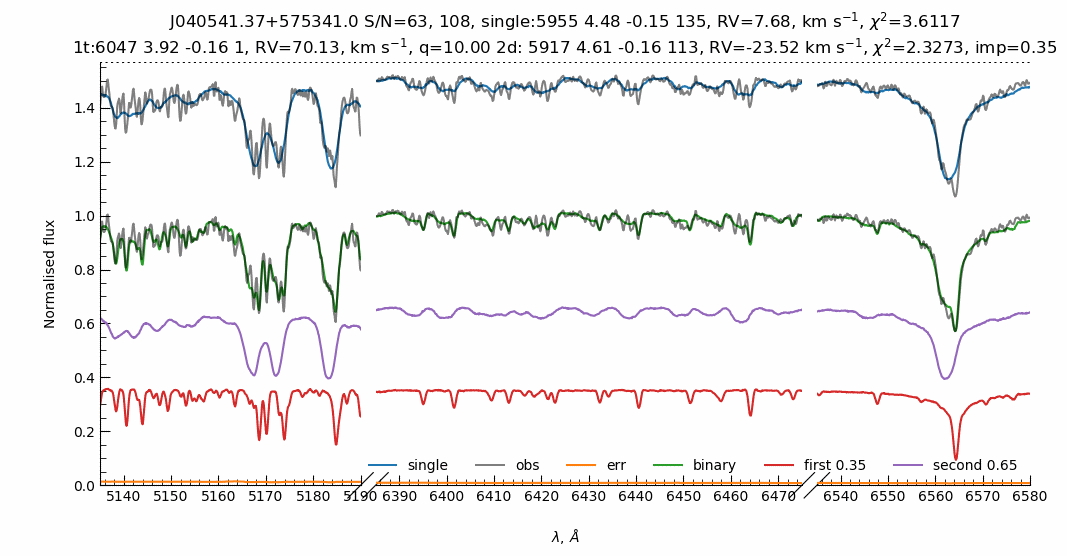}
	\caption{Same as Fig.\protect\ref{fig:faster} but for an SB3 candidate.}
    \label{fig:sb3}
\end{figure*}

Fig.~\ref{fig:sb3} shows a spectrum with three spectral components (SB3 J040541.37+575341.0), where the primary component of the binary model broadens to cover two out of three components. In \citet{m11} we showed how such a spectrum can be disentangled. 

\begin{figure*}
	\includegraphics[width=0.83\textwidth]{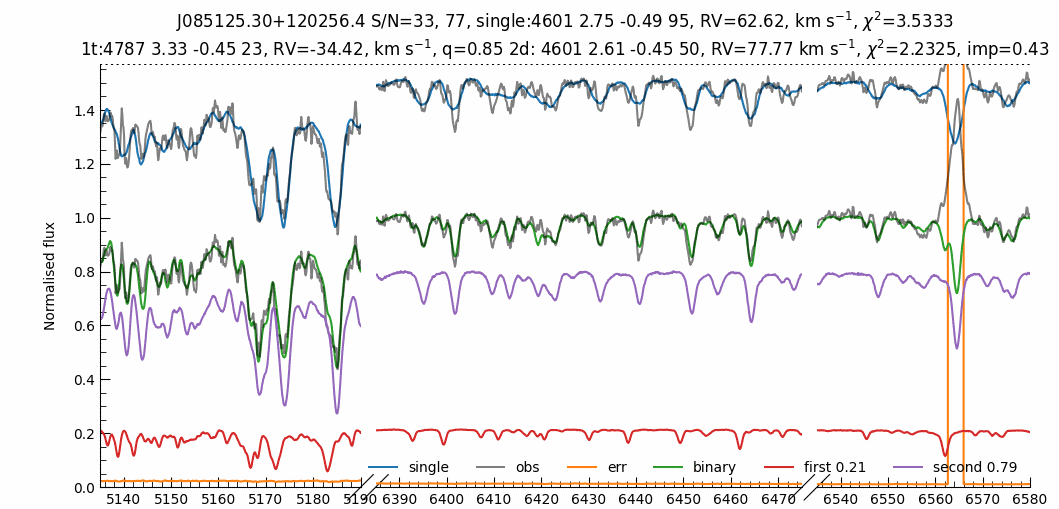}
	\includegraphics[width=0.83\textwidth]{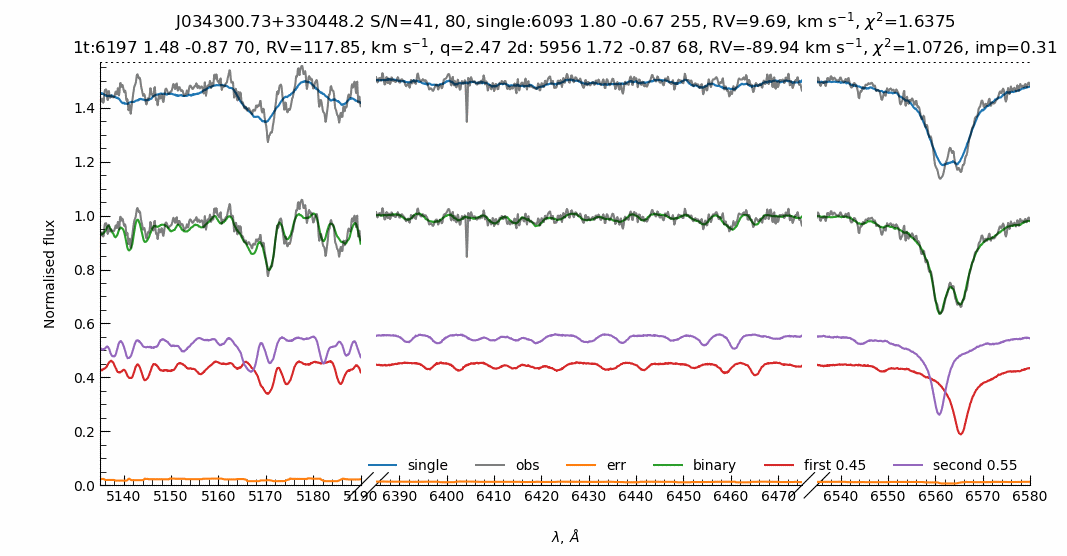}
	\caption{Same as Fig.\protect\ref{fig:faster} but for SB2s with matches in SB1 (top panel) and a single star (bottom panel) tables from \protect\citet{bardy2018}}.
    \label{fig:elbadry}
\end{figure*}

In Fig.~\ref{fig:elbadry} we show the best fit single star and binary models for two SB2s, which are reported as SB1 (J085125.30+1202564) and single star (J034300.73+330448.2) in \citet{bardy2018}. Double lines are clearly seen in the spectra, however they still can be chance alignments or the LAMOST-MRS was lucky to observe them at optimal $\drv$.  

\subsubsection{non-stacked spectra}
\label{sec:subepochs}

\begin{figure*}
    \includegraphics[width=0.83\textwidth]{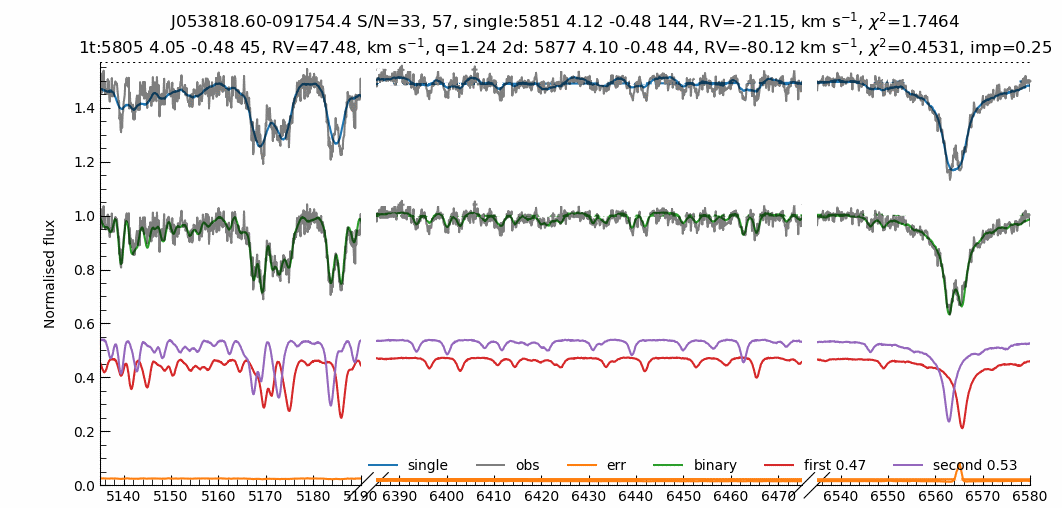}
	\includegraphics[width=0.83\textwidth]{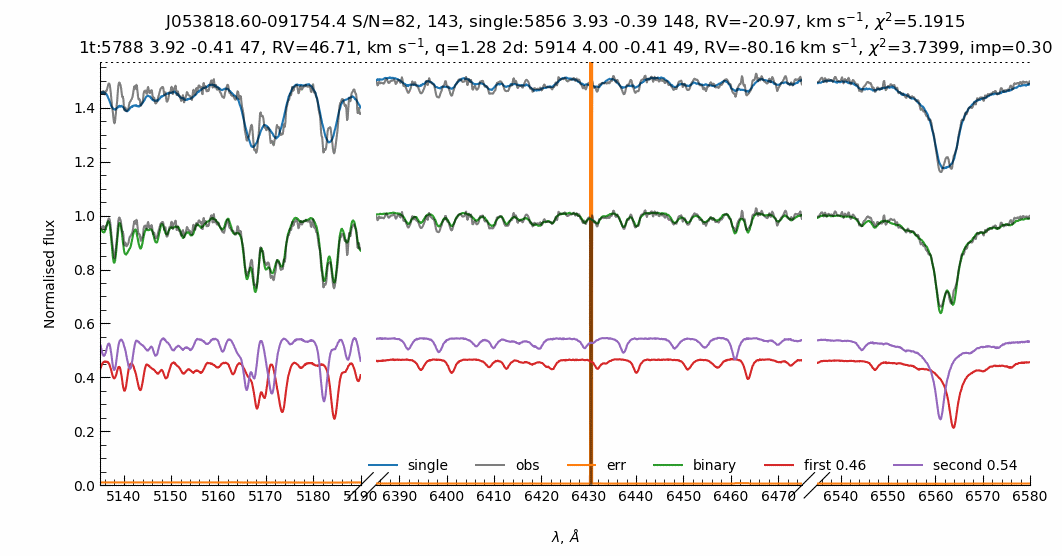}
	\includegraphics[width=0.83\textwidth]{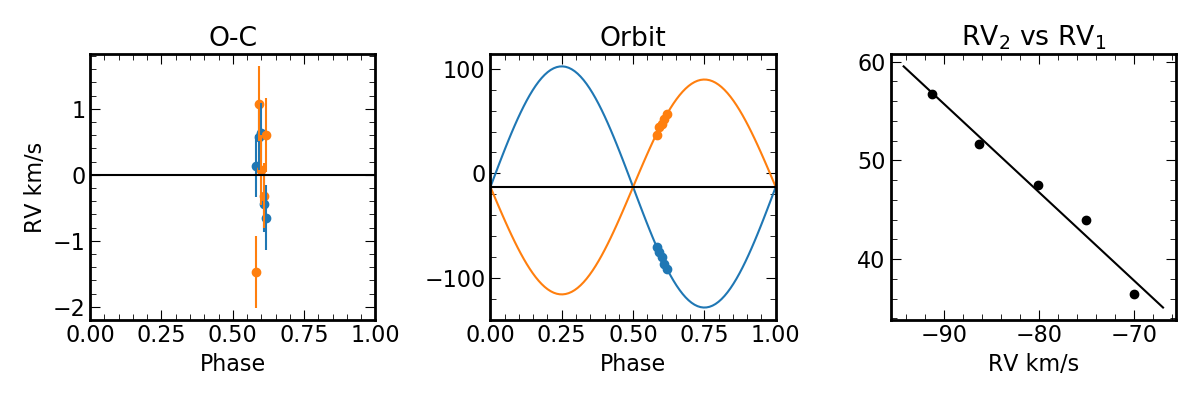}
	\caption{Same as Fig.\protect\ref{fig:faster} for non-stacked (top panel), stacked spectrum (middle panel) of close SB2. In the bottom panels we show a circular orbit fitting with the phase-folded $\rv$ curve in the middle, the fit residuals on the left and the Wilson plot on the right.}
    \label{fig:orbit}
\end{figure*}

In Fig.~\ref{fig:orbit} we show the best fits for the non-stacked and the stacked spectra of the close binary J053818.60-091754.4. We successfully fit five individual 20~min exposures with lower $\snr_{\rm blue,red}\sim30,~60$ (stacked spectrum has $\snr_{\rm blue,red}\sim82,~143$). We find that $\rv$ rapidly changes and $\vsini_{1,2}$ are slightly smaller ($\sim3-5\kms$) than for the stacked spectrum. Similarly to \citet{m11} we fit a circular orbit with period $P=1.88$ days, mass ratio $q_{\rm dyn}=0.89$ and systemic velocity $\gamma=-13.14~\kms$, although orbit coverage is very poor.  

\subsubsection{not detected SB2s}
\label{sec:miss}
\begin{figure*}
	\includegraphics[width=0.83\textwidth]{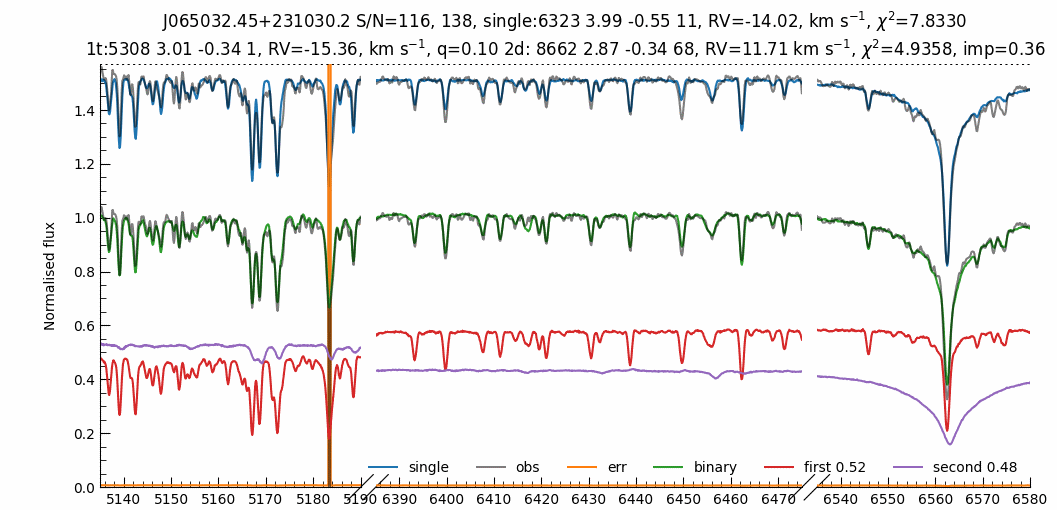}
	\includegraphics[width=0.83\textwidth]{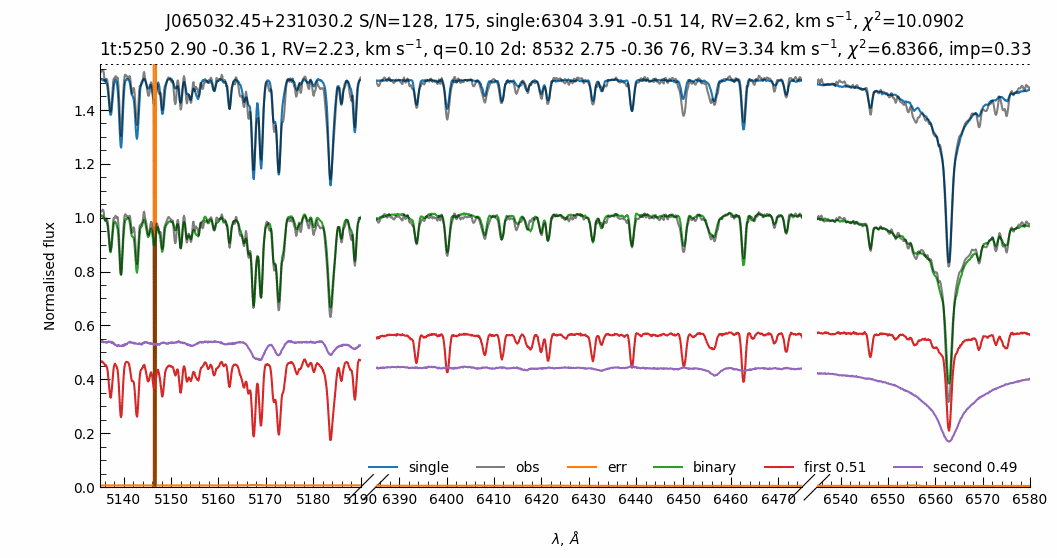}
	\caption{Same as Fig.\protect\ref{fig:faster} but for two spectra of SB2 system with two very different components with $\Delta\teff\sim3000$~K. Top panel shows the spectrum with $\drv=27~\kms$, and bottom panel shows the spectrum with $\drv\sim0~\kms$.  }
    \label{fig:gig}
\end{figure*}

In Fig.~\ref{fig:gig} we show the best fit single star and binary models for J065032.45+231030.2 -  a very interesting SB2 system of two component with $\Delta\teff\sim3000$~K, denoted as ``gig" in Fig.~\ref{fig:select}. It is not selected by our criteria due to a fast rotating secondary $\vsini=68~\kms$, however it can be selected based on high $\imp>0.3$ even for a spectrum with $\drv\sim0~\kms$. We will study this SB2 in detail in Kovalev et al. (in prep.)

\begin{figure*}
	\includegraphics[width=0.83\textwidth]{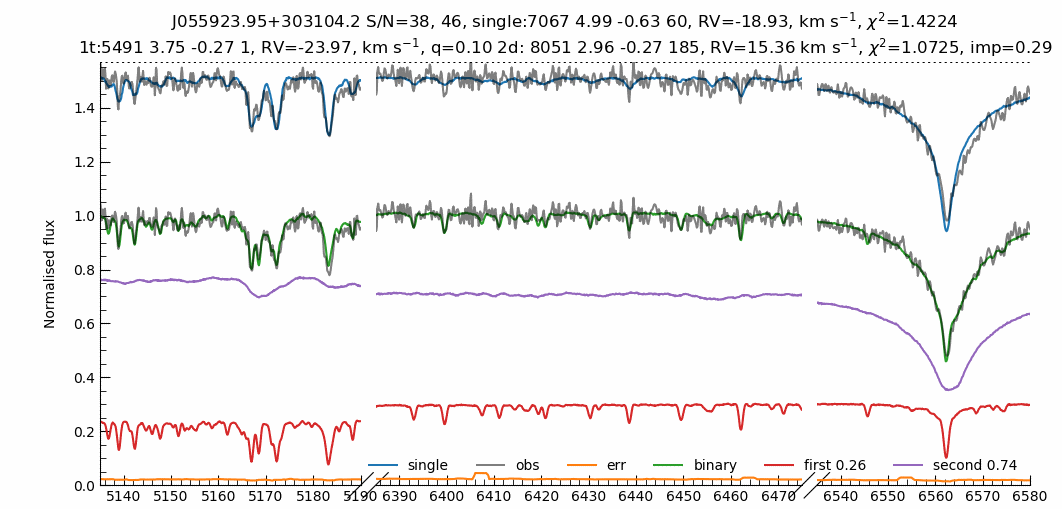}
	\includegraphics[width=0.83\textwidth]{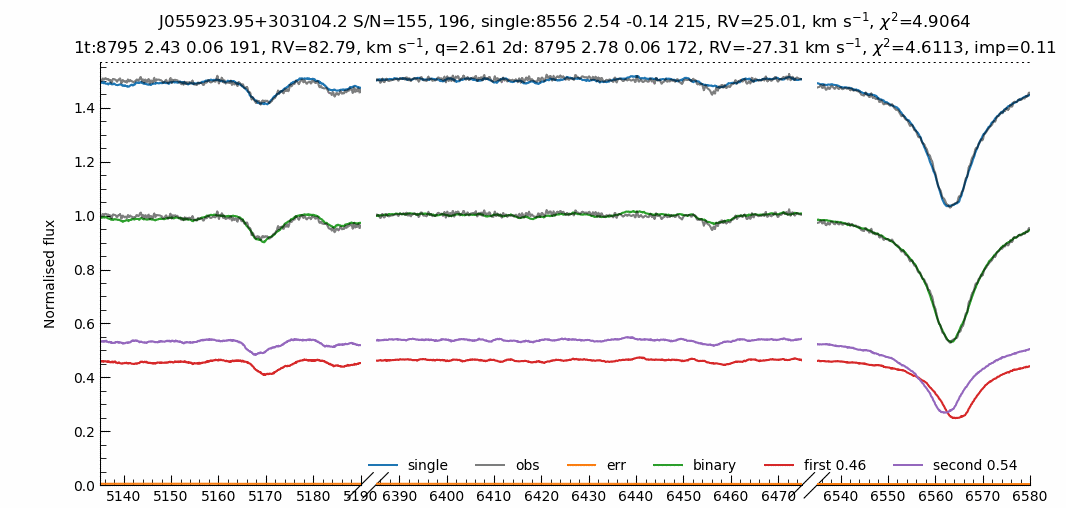}
	\includegraphics[width=0.83\textwidth]{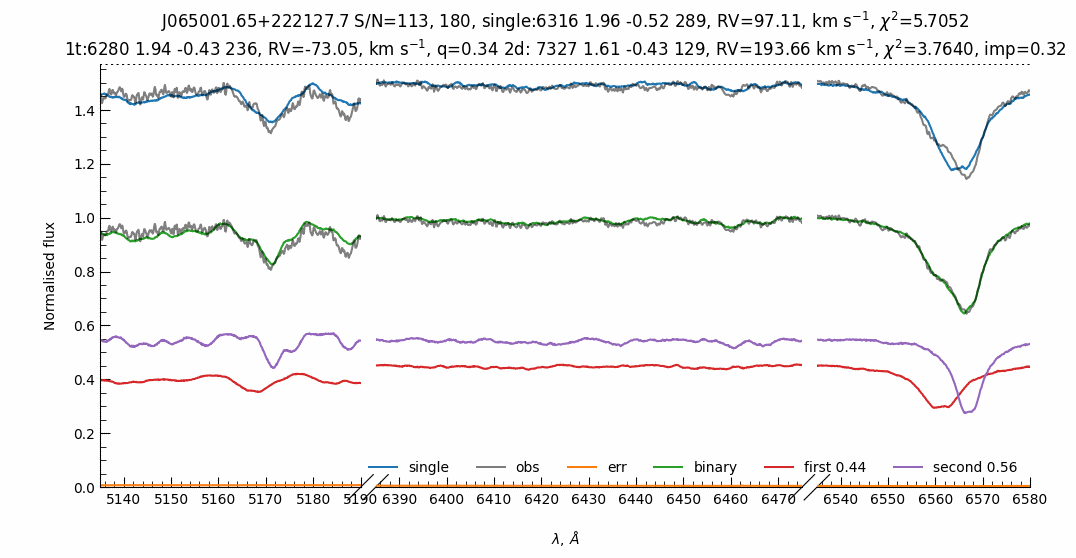}
	\caption{Same as Fig.\protect\ref{fig:faster} but for two SB2 systems not selected by our method. Top panel shows a spectrum, possibly taken during partial eclipse, where single star model ignores the primary component, but fits only narrow lines of the secondary. Middle panel shows the spectrum of the same star, but without narrow lines. The bottom panel shows the spectrum of an SB2 system of two very fast rotators with insufficient $\drv$.  }
    \label{fig:miss}
\end{figure*}

In Fig.~\ref{fig:miss} we show two SB2 systems not selected by our method. The spectrum of J055923.95+303104.2 shown in the top panel is a case~\ref{case3}, where single star model fits only narrow lines of the secondary. The middle panel shows the same star without narrow lines of the secondary and the binary model consists two hot, fast rotators which is probably wrong, as their $\teff_{1,2}$ lie at edge of the model grid. The bottom panel shows the spectrum of the SB2 system J065001.65+222127.7 of two very fast rotators with insufficient $\drv$.  We will study these SB2s in detail in Kovalev et al. (in prep.)

\subsection{Bad fits}

\begin{figure*}
	\includegraphics[width=0.83\textwidth]{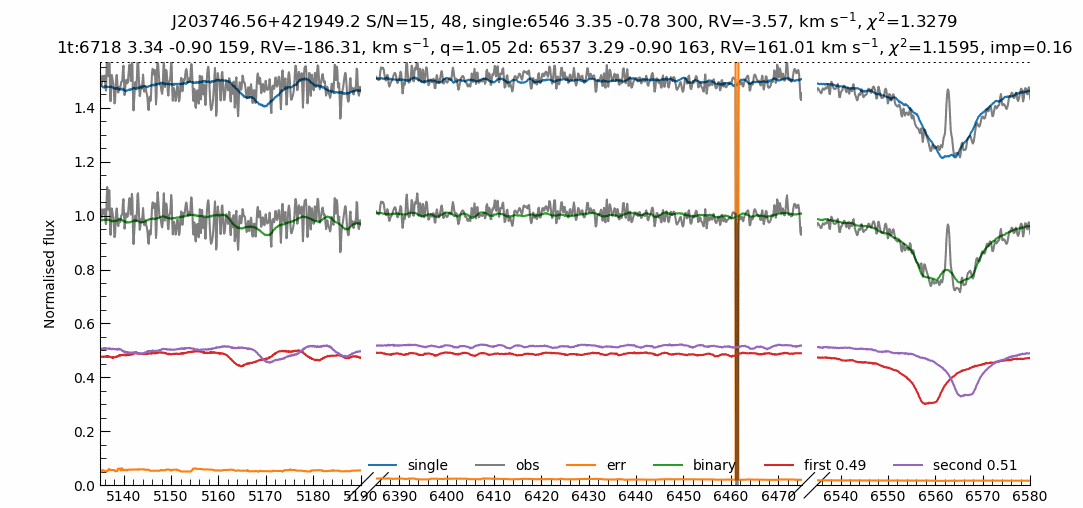}
	\includegraphics[width=0.83\textwidth]{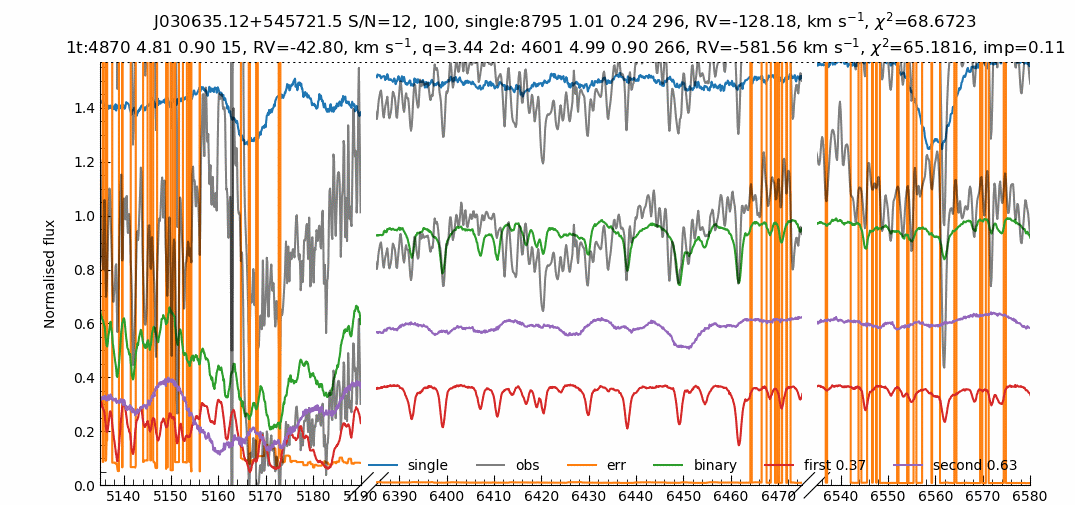}
	\caption{Same as Fig.\protect\ref{fig:faster} but for poor fit of a very hot star J203746.56+421949.2 (top panel) and a very cool star J030635.12+545721.5 (bottom panel)}
    \label{fig:bad_fits}
\end{figure*}

In Fig.~\ref{fig:bad_fits} we show how our model desperately attempts to fit a very hot star J203746.56+421949.2 (top panel) and a very cool star J030635.12+545721.5 (bottom panel). For a hot star with emission at the $\ha$ core, our binary model fits ``fake" system with $\drv\sim350~\kms$ and good $\imp$. For the spectrum of a cool star with clearly visible molecular bands, both models failed, but the secondary component in the binary model was able to estimate $\rv$ correctly. 

\subsection{Bad wavelength calibration}
\label{sec:bads}
\begin{figure*}
	\includegraphics[width=0.83\textwidth]{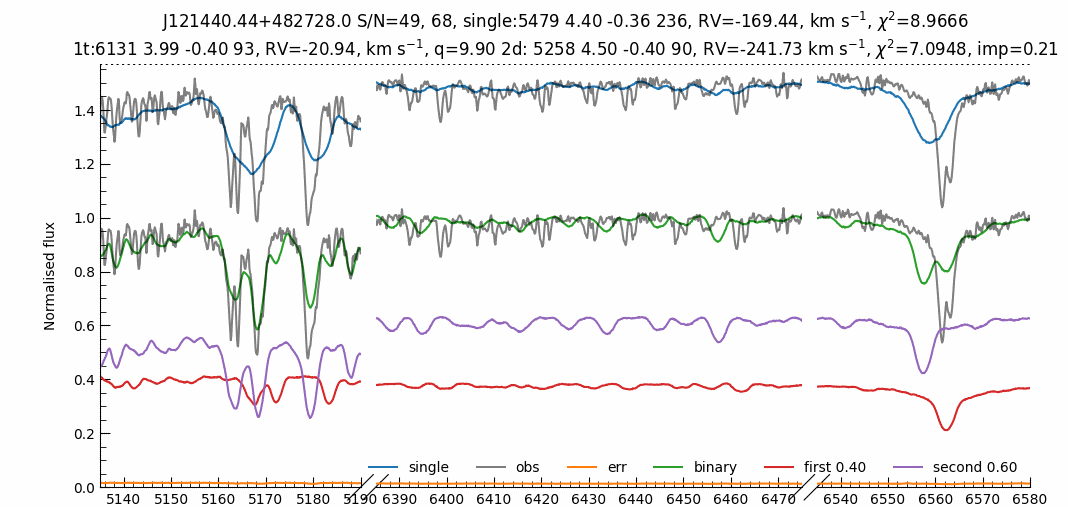}
	\includegraphics[width=0.83\textwidth]{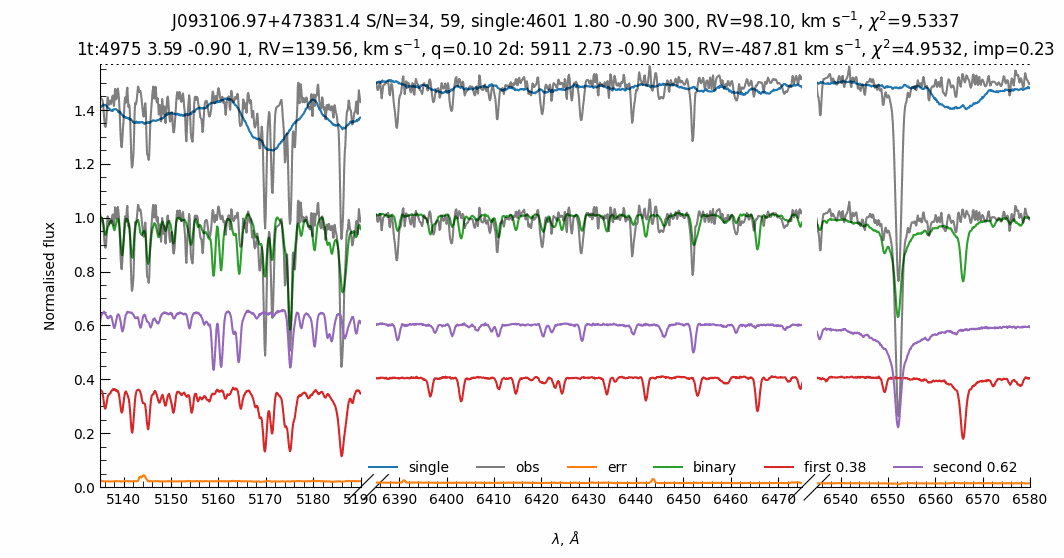}
	\caption{Same as Fig.\protect\ref{fig:faster} but for spectra with bad wavelength calibration. Top panel is a SB2 candidate J121440.44+482728.0 and bottom panel is a single star J093106.97+473831.4.}
    \label{fig:bad_lambda}
\end{figure*}

We identify 213 spectra, where the blue and red spectral arms have nearly constant $\drv>180~\kms$, after visual inspection of fits. All these spectra were observed in several particular fields and nights, thus we assume that the wavelength calibration was bad for certain fibers of the spectrograph. We list such spectra in Table~\ref{tab:calibr}. Examples of such spectra of the SB2 (top panel) and single star (bottom panel) are shown in Fig.~\ref{fig:bad_lambda}. You can see that for a single star J093106.97+473831.4 the single star model completely failed, but the binary model was able to fit $\rv$ from the red arm by primary component and $\rv$ from the blue arm by the secondary component. For the SB2 star J121440.44+482728.0 even the binary model failed, but showed a result similar to the single star model fits on normal SB2 spectra: both components are broadened.

\begin{table}
    \centering
    \caption{Spectra with a problem in the wavelength calibration. Full table is available as supplementary material.}
    \begin{tabular}{ccccc}
\hline
\hline
star & RV$_{0}$ & RV$_{2}$&RV$_{1}$& MJD   \\
  & $\kms$& $\kms$ & $\kms$& d\\

\hline
J010308.47+053110.8& 232.6 & -2.3 & 328.4 & 58790.635\\
J010355.51+052339.3& 230.7 & 26.4 & 356.1 & 58790.635\\
 ..& ..& .. &  ..& ..\\

\hline
    \end{tabular}
    \label{tab:calibr}
\end{table}

\section{Catalogue of SB2 candidates in LAMOST-MRS}

Table~\ref{tab:final} lists all 2460 SB2 candidates. We note that LAMOST designation can slightly vary between data releases, therefore please use Gaia eDR3 \texttt{source\_id} from \citet{egdr3}.  

\begin{table}
    \centering
    \caption{Catalogue of SB2 candidates in LAMOST-MRS. Full table is available as supplementary material.}
    \begin{tabular}{cc}
\hline
\hline
LAMOST designation & Gaia eDR3 \texttt{source\_id} \\
\hline
J000024.99+340651.5 & 2875121652781859200\\
J000151.03+340757.3 & 2875206688837760640\\
J000223.56+340644.8 & 2875159070536814464\\
..&..\\

\hline
    \end{tabular}
    \label{tab:final}
\end{table}

\section{New selection in M~11 }
\label{sec:list_m11}

\begin{figure*}
	\includegraphics[width=0.83\textwidth]{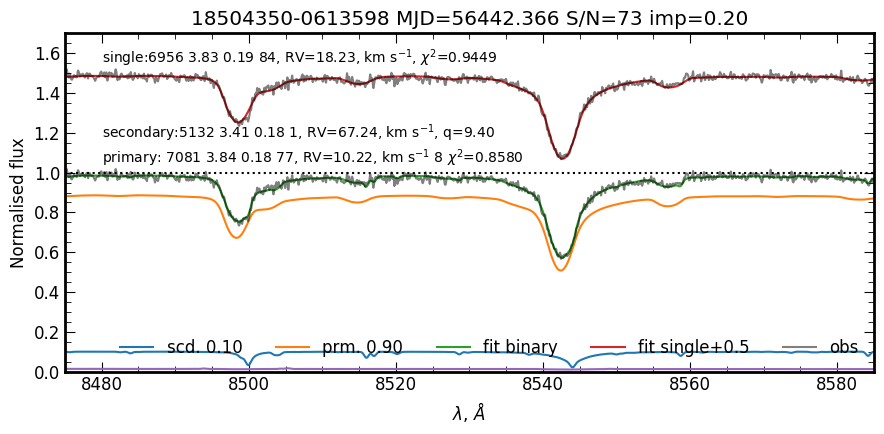}
	\includegraphics[width=0.83\textwidth]{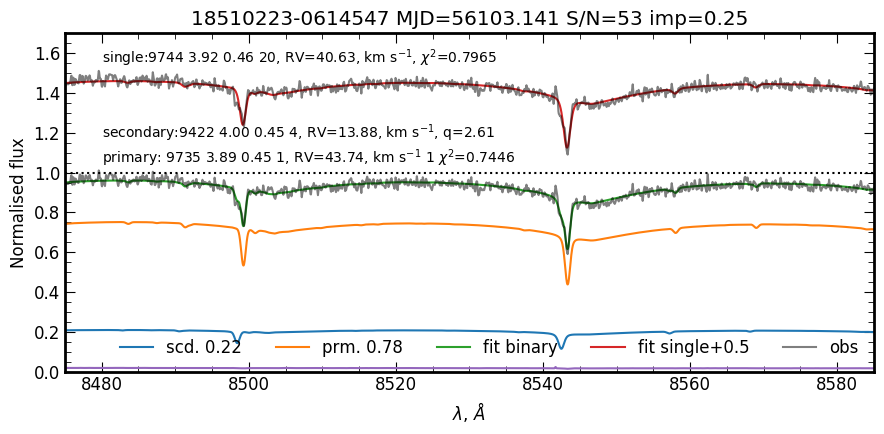}
	\caption{New (top panel) and not selected (bottom panel) SB2 candidates in M~11. }
    \label{fig:3m11}
\end{figure*}

The new SB2 candidate is 18504350-0613598 shown on the top panel of Fig.\ref{fig:3m11}. The very high mass ratio $q=9.4$ and small $\logg_2=3.41$~dex are not realistic, probably the binary model converges to such parameters to make better fit of the light ratio. SB2 candidate 18510223-0614547 from \cite{merle2017,m11} shows clear composite spectrum (bottom panel of Fig.\ref{fig:3m11}), although it is not selected due to $|\rv_1-\rv_0|<5~\kms$ (case~\ref{case3}). All other six SB2 and one SB3 candidate from \cite{m11} with $\vsini_1+\vsini_2+5<\vsini_0$ are selected.  
 Table~\ref{tab:cat_m11} lists stellar parameters derived in \cite{m11} for all 265 spectra in M~11.     
Plots with all 265 spectral fits will be available online \url{https://doi.org/10.5281/zenodo.7037605}\footnote{also available at \url{https://nlte.mpia.de/upload/kovalev/}}.

\begin{table}
    \centering
    \caption{Stellar parameters for all 265 M~11 stars, analysed in \protect\cite{m11}. Full table in machine-readable form is available  as supplementary material. }
    \begin{tabular}{cc}
\hline
\hline

parameter & unit \\

\hline
CNAME & HHMMSSss+DDMMSSs\\
MJD & day\\
$\chi^2_{\rm single}$ & \\
$\chi^2_{\rm binary}$ & \\
$\imp$ & \\
$\alpha$ (J2000) & degree\\
$\delta$ (J2000) & degree\\
$\snr$ & pix$^{-1}$\\
$\rv_0$ & $\kms$\\
$\rv_2$ & $\kms$\\
$\rv_1$ & $\kms$\\
$q$ & \\
${\rm frac}=k/(1+k)$ & \\
$\teff_0$ & K\\
$\logg_0$ & dex\\
$\feh_0$ & dex\\
$\vsini_0$ & $\kms$\\
$\teff_2$ & K\\
$\logg_2$ & dex\\
$\feh_{1,2}$ & dex\\
$\vsini_2$ & $\kms$\\
$\teff_1$ & K\\
$\logg_1$ & dex\\
$\vsini_1$ & $\kms$\\
  Gaia eDR3 \texttt{source\_id} & \\
\hline
    \end{tabular}
    \label{tab:cat_m11}
\end{table}


\bsp	
\label{lastpage}
\end{document}